\documentclass{article}
\usepackage[english]{babel}
\usepackage[T2A]{fontenc}
\usepackage[cp1251]{inputenc}
\usepackage[dvips]{graphicx}
\usepackage{amsmath}
\usepackage{amssymb}
\usepackage{latexsym}
\begin{document}
\rm
\let\sc=\sf
\begin{flushright}
Journal-Ref: Astronomy Letters, 2010, Vol. 36, No. 7, pp. 498-505
\end{flushright}
\begin{center}
\LARGE {\bf Brightness Oscillations in Models of Young Binary Systems
with Low-Mass Secondary Components}\\

\vspace{1cm}
\Large {\bf T. V.\,Demidova$^{1,2}$,
V. P.\,Grinin$^{1,2}$, N. Ya.\,Sotnikova$^2$}

\normalsize \vspace{5mm}
1 - Pulkovo Astronomical Observatory, Russian Academy of Sciences, Pulkovskoe shosse 65, St. Petersburg,
                                         196140 Russia, \\
2 - Sobolev Astronomical Institute, St. Petersburg State University, Universitetskii pr. 28, St. Petersburg, 198504 Russia,
\\

\end{center}
Received September 7, 2009
\normalsize
\begin{abstract}

We consider a model for the cyclic brightness variations of a young star with a low-mass
companion that accretes matter from the remnants of a protostellar cloud. At small inclinations of the
binary orbit to the line of sight, the streams of matter and the density waves excited in the circumbinary
disk can screen the primary component of the binary from the observer. To study these phenomena, we
have computed grids of hydrodynamic models for binary systems by the SPH method based on which we
have constructed the phase light curves as a function of the rotation angle of the apsidal line relative to the
observer. The model parameters were varied within the following ranges: the component mass ratio $q =
0.01-0.1$ and the eccentricity $e = 0-0.5$. We adopted optical grain characteristics typical of circumstellar
dust. Our computations have shown that the brightness oscillations with orbital phase can have a complex
structure. The amplitudes and shapes of the light curves depend strongly on the inclination of the binary
orbit and its orientation relative to the observer and on the accretion rate. The results of our computations
are used to analyze the cyclic activity of UX Ori stars.\\

Key words: \emph{young binary systems, hydrodynamics, cyclic activity, accretion}.
\end{abstract}
\clearpage
\clearpage
\large
\newpage

\section{INTRODUCTION}
In this paper, we continue our numerical simulations
of the cyclic circumstellar extinction variations
in young binary systems accreting matter from the
remains of a protostellar clouds begun in our previous
works (Sotnikova and Grinin 2007; Demidova et al.
2010). The goal of these computations is to as certain
what photometric effects may be expected when the
binary orbit is inclined at a small angle to the line
of sight. Sotnikova and Grinin (2007) showed that
three harmonics could be observed in the circumstellar
extinction variations in such binaries as a result
of periodic gravitational perturbations produced by
the orbital motion of the components. The shortest
of them has a period equal to the orbital one and is
produced by the streams of matter that periodically
appear at the inner boundary of the circumbinary
disk and penetrate into its central region. The other
two, longer periods are attributable to the precession
of the circumbinary disk and the motion of a one armed
density wave in it. All three harmonics can
be observed simultaneously only in the presence of
sufficiently strong perturbations. Such conditions can
take place in binary systems whose components do
not differ greatly in mass (this case was considered by
Demidova et al. (2010)). In this paper, we consider binary
systems with low-mass secondary components
in which only one mode of extinction oscillations with
a period equal to the orbital one is realized. We use
a binary model that consists of a primary component
with mass $M_1$ and a secondary component with
mass $M_2$. The binary is embedded in a circumbinary
gas-dust disk whose matter is accreted onto the
binary components (Fig. 1). The disk is assumed to
be coplanar with the orbit. The extreme case of such a
binary is a young star with a protoplanetary disk and
a giant planet at the phase of intense accretion. The
input parameters of the problem are the accretion rate
onto the binary components ($\dot M_a$), the component
mass ratio $q = M_2/M_1$, the orbital inclination to the
line of sight $\theta$, the eccentricity $e$, the rotation angle
of the apsidal line relative to the observer $\phi$, and the
parameter $c$ characterizing viscosity. The goal of this
paper is to compute the light curves of such binaries
and to study their dependence on model parameters.
\section{THE COMPUTATIONAL METHOD}
As in our previous papers, we consider a binary
system that accretes matter from a circumbinary
(CB) disk coplanar with the orbital plane. The hydrodynamic
models of such a binary were computed
by the SPH (Smoothed Particle Hydrodynamics)
method described in detail by Sotnikova (1996). For
each model, we computed the column density of test
particles toward the primary component of the binary
as a function of time expressed in units of the orbital
period. As a rule, the computations were performed
for several hundred periods.

The area of the section of the column $s$ along
which the particle column density in the chosen direction
was determined was specified as $s = 2h \times 2h$.
In our implementation of the method, the smoothing
length was assumed to be constant and was specified
in fractions of the semimajor axis of the binary
orbit, $h =0.1a$. At each point, this provided at least
$30$ neighboring points over which the hydrodynamic
quantities were averaged. Our computations showed
this value of $s$ to be optimal for the solution of the
formulated problem (at lower values of s, the influence
of fluctuations is enhanced; at higher values, the time
resolution of the calculated characteristics deteriorates).

Our quantitative analysis of the computational results
began with the removal of the trend in the variations
of the test particle column density attributable
to the decrease in their number in the binary system
because of accretion onto its components. The
trend was modeled by a fifth-degree polynomial. Our
computations showed that this provided a satisfactory
removal of the trend for all of the models considered.
Subsequently, we passed from the column
density of test particles $n(t)$ to the column density
of real dust grains $n_d (t)$ (for more detail, see Demidova
et al. 2010). To reduce the influence of random
fluctuations when the phase dependences of $n$ were
computed, the current values of $n(t)$ were folded with
the orbital period for a time interval of $50$ binary
revolutions.

One of the key parameters of the computed models
is the specified accretion rate onto the binary components
$\dot M_a$. This parameter was compared with the
accretion rate of test particles obtained during the
computations (for more detail, see Sotnikova and
Grinin 2007). The mass of a single test particle was
determined as follows: $m_{d}= P \cdot \dot M_a/N$. Here, $P$ is the
orbital period and $N$ is the total number of test particles
accreting onto both components in one binary
revolution. Below, in our computations, we took $\dot M_a$
to be $10^{-9}\, M_{\odot}/yr$. As our computations showed,
this value of the accretion rate is high enough to
produce a strong modulation in the brightness of the
primary component.

Having calculated the mass of a single test particle
by the method described above, we obtain the matter column density in $g/cm^2$. To determine the
optical depth of the dust on the line of sight, we
should specify the opacity $\kappa$ per gram of matter. This
parameter depends on the type and sizes of the dust
grains and on the dust-to-gas ratio. As in our previous
papers, below we adopted an average (for the
interstellar medium) dust-to-gas ratio of $1 : 100$ and
$\kappa = 250\,cm^{2}/g$ typical of circumstellar extinction
in Johnson's $B$ photometric band (Natta and Whitney
2000)\footnote{It should be noted that these circumstellar dust parameters
correspond to early evolutionary stages of protoplanetary
disks. In the process of coagulation, the dust grains
become larger and settle toward the disk plane. However,
as observations show, the small dust grains that make a
major contribution to the opacity of matter are retained in the
protoplanetary disks for a long time, of the order of several
Myr. As the calculations by Birnstiel et al. (2009) show, this
is because the efficiency the process opposite to coagulation,
the destruction of particles in collisions, is high.}. Multiplying the matter column densities
calculated in this way by $\kappa$, we will obtain the optical
depths $\tau$ for each instant of time.

The intensity of the radiation from young stars is
known to consist of two components: the intensity of
the direct stellar radiation $I_{\ast}$ (in our case, the primary
component of the binary) attenuated by a factor of
$e^{- \tau}$ and the intensity of the radiation scattered by
circumstellar dust $I_{sc}$:

\begin{equation}
I_{obs}=I_{\ast}e^{-\tau} + I_{sc}\,, \label{Iobs}
\end{equation}

The contribution of the scattered light to the total
radiation typically does not exceed a few percent.
Therefore, below, when studying the pattern of variability
in the primary component of the binary, we
took the intensity of the scattered radiation in (1) to
be zero. The light variations of the primary component
are expressed in magnitudes: $\Delta m =
-2.5\cdot \log{I_{obs}}$ ($I_{\ast}$ is taken as unity). Hence it follows that $\Delta m \sim \tau$, and
since $\tau$ is proportional to $\dot M_a$,

\begin{equation}
\Delta m \sim \dot M_a
\end{equation}

This relation allows the light curves computed for one
value of $\dot M_a$ to be recalculated for other values of this
parameter.

\section{MODEL LIGHT CURVES}
We computed theoretical light curves for binary
systems by the method described above. The basic
model parameters are listed in the table. In all models,
the mass of the primary component is $2M_\odot$, i.e., a
value typical of UX Ori stars (Rostopchina 1999);
$c$ is the dimensionless speed of sound in the matter
expressed in units of the orbital velocity of the secondary
component at $e = 0$. The parameter $c$ enters
into the expression for the disk viscosity. Note that the
values of $c$ used in our hydrodynamic computations
correspond to a matter temperature of the order of
several hundred kelvins. The number of test particles
$N$ used in our SPH simulations is 60 000 and the
smoothing length is $h = 0.1a$. As in Demidova et al.
(2010), the orbital period is taken to be five years.

The model parameters were varied within the
following ranges: the component mass ratio $q = M_2/M_1 = 0.01-0.1$ and the eccentricity $e = 0-0.5$;
the dimensionless speed of sound in the CB disk $c$ was
taken to be $0.05$ ("warm" disk) for all of the models
except model 5. For this model, $c = 0.02$ ("cold"
disk). Our computations were performed for several
inclinations of the binary orbit to the line of sight
and four rotation angles of the apsidal line relative to
the observer: $0^\circ$, $90^\circ$, $180^\circ$, and $270^\circ$. In the frame of
reference adopted here, the angle $\phi = 0$ corresponds
to the case where the apastron lies between the
primary component and the observer. The angles
were counted off in the direction of rotation of the
binary. The choice of inclination (from $0^\circ$ to $12^\circ$) was
restricted by the finite number of test particles used
in our computations and by the necessity of avoiding
great statistical fluctuations in the particle column
density at larger angles $\theta$. As was said above, to
suppress the fluctuations, the current values of $n(t)$
for the chosen phase intervals were summed over
50 revolutions and were then averaged and smoothed.
As an example, Fig. 2 shows the light curve computed
in this way along with the "cloud" of points from
which it was obtained.

Figures 3-6 present the light curves of a binary
system in models with eccentric orbits for two indications
of the disk plane to the line of sight and
several rotation angles of the apsidal line relative to
the observer. We see that the light curves depend
significantly not only on the orbital inclination to the
line of sight but also on the orientation of the apsidal
line relative to the observer. Both the amplitude and
the shape of the light curves change with these parameters.

The dependence of the light-curve shape on the
orbital inclination to the line of sight is quite understandable
if we take into account the fact that the
extinction variations with orbital phase are caused
partly by the streams of matter propagating from the
CB disk to the central part of the binary system and
partly by the CB disk matter falling on the line of sight
as it rotates. The strong influence of the orientation of the apsidal line relative to the observer on the light
curves in models with eccentric orbits is caused by an
azimuthal asymmetry of the inner CB disk gap with a
nearly elliptical shape.

In model 1 (Fig. 3) with the component mass ratio
$q = 0.1$ and eccentricity $e = 0.3$ at a given accretion
rate, an appreciable variability amplitude is observed
for all four orientations of the binary orbit relative to
the observer. At rotation angles of the apsidal line
relative to the observer of $0^\circ$ and $180^\circ$, the amplitudes
for inclinations of $0^\circ$ and $7^{\circ}.5$ are comparable, because
the size of the column section is finite: at such sizes,
the number of test particles in the column was found
to be close in order of magnitude at a difference in
angles of $7^{\circ}.5$.

In model 2 (Fig. 4), we considered inclinations
of $0^{\circ}$ and $12^{\circ}$. For the inclination of $0^{\circ}$, the minimum
variability amplitude was obtained for rotation
angles of the apsidal line of $0^{\circ}$ and $90^{\circ}$, while for
the inclination of $12^{\circ}$ the amplitude is at a minimum
for $180^{\circ}$ and $270^{\circ}$. This suggests that the CB disk
is azimuthally very inhomogeneous even when the
secondary component is a factor of $10$ less massive
than the primary component. It is this inhomogeneity
that is reflected in the behavior of the particle column
density.

In models 3 and 4 (Figs. 5 and 6), we adopted
the minimum masses of the secondary component
we considered, $q = 0.03$ and $q = 0.01$. The variability
amplitude for an inclination of $0$ is $2-3^m$; for inclinations
of $9^\circ$ and $10^\circ$, it is about $0^{m}.8$. Thus, the
noticeable photometric variability of the star may be
due to the orbital motion of a brown dwarf or a giant
planet. For model 4 (with a circular orbit), the result of
our computations does not depend on the orientation
of the binary orbit.

In model 5 (with the lowest viscosity), the disk is
too thin and dense. As a result, when observed edge on,
the radiation from the primary component is completely
screened by the disk, while for an inclination of
$5^\circ$ the number of particles on the line of sight does not
exceed the fluctuation level.

The above-listed results were obtained without allowance
for the influence of scattered radiation. Nevertheless,
they are also valid in real systems with a noticeable
contribution from scattered light. This can be
seen from Fig. 7, which shows two light curves. One
of them was computed for  $I_{sc} = 0$ and the other was
computed for $I_{sc} = 0.1I_{\ast}$. We see that the scattered
radiation reduces the amplitude of the light variations,
but their overall appearance is retained.

It should be noted that to obtain statistically significant
results in counting the number of test particles
on the line of sight, we had to restrict ourselves
to a small range of orbital inclinations to the line of
sight ($\le 12^\circ$). To reduce the influence of fluctuations
in SPH models, it is necessary to use a larger number
of particles, which requires much computational time.
Test computations showed that in such models an
appreciable brightness modulation amplitude could
also be obtained at larger inclinations of the binary
orbit to the line of sight (especially in the models with a "warm" and "hot" CB disk), but this requires higher
accretion rates. Note that the accretion rate adopted
in our computations ($10^{-9}M_\odot$) is the lower limit
of $\dot M_a$ for young stars. The accretion rate onto Herbig
Ae/Be stars is known (see, e.g., Garcia Lopez et al.
2006) to reach $10^{-6} M\odot$. This means that, in
our case, there is a large margin (about three orders
of magnitude) for increasing this parameter.

\begin{table}[p]
\label{t_models}
\begin{center}
\caption{Model parameters}
\begin{tabular}{c|c|c|c|}
\hline \hline
  Model & $e$ &  $q$ & $c$ \\
\hline
  1 & 0.3 & 0.1 & 0.05 \\
  2 & 0.5 & 0.1 & 0.05 \\
  3 & 0.5 & 0.03& 0.05 \\
  4 & 0.0 & 0.01& 0.05 \\
  5 & 0.5 & 0.1& 0.02  \\
\hline
\end{tabular}
\end{center}
\end{table}

\section{DISCUSSION AND CONCLUSION}

The above results are quite expectable, because it
is clear from general considerations that something
must vary with a period equal to the orbital one in
young binary systems that accrete matter from the
remnants of a protostellar cloud. The unexpected and
nontrivial result of our computations is the conclusion
that the source of perturbations capable of generating
a photometric wave with an appreciable amplitude in
the behavior of the brightness of a young star could
be a companion that is a factor of $100$ less massive
than the star itself. This means that a brown dwarf or
a giant planet can be such a companion. Therefore,
studying the cyclic activity of young stars can contribute
to the discovery and study of such objects.

Our computations showed that an accretion rate
onto the binary components of $10^{-9}\, M_{\odot}$  is
quite sufficient to produce a strong modulation in the
brightness of the primary component at the orbital
inclinations for which the computations were performed.
It means that at a "favorable" (nearly edge on)
disk orientation, the cyclic activity caused by
extinction variations can be observed even in young
objects with weak evidence of accretion. An example
of such an object is the WTTS star (weak-line T Tauri
star) V718 Per, in which shallow eclipses with a
period of 4.7 yr and a duration of 3.5 yr are observed
(see, e.g., Grinin et al. 2008).

The accretion rate onto UX Ori stars to be
discussed below is approximately an order of magnitude
higher (Tambovtseva et al. 2001; Muzerolle
et al. 2004). This means that the cyclic activity of such
stars caused by periodic extinction variations can be
observed at larger inclinations of the circumstellar
disks to the line of sight. In models with eccentric
orbits, the amplitude and shape of the phase light
curves depend not only on the inclination of the binary
orbit to the line of sight but also on its orientation
in space; as was noted above, this is attributable
to the elliptical shape of the inner gap and can be
observed even in binaries with a low-mass secondary
companion. At certain orbital inclinations, both the
CB disk matter and the streams of matter penetrating
into the inner regions of the binary are involved in
producing variable extinction on the line of sight. In
such cases, the light curves can have a fairly complex
shape (see, e.g., Fig. 5).

Sotnikova and Grinin (2007) and Demidova et al.
(2010) pointed out the possible connection of the
cyclic activity of UX Ori stars with the extinction
variations due to the presence of perturbing bodies
(protoplanets, brown dwarfs, components of binary
systems) in their neighborhoods. Observations show
that the duration of the photometric cycles ranges
from several months (Artemenko et al. 2010) to ten
or more years (see, e.g., Shakhovskoi et al., 2005).
Their amplitudes range from a few tenths of a magnitude
to two magnitudes in the $V$ band, i.e., they are
comparable to the theoretical values.

As an example, Figure 8 shows the phase light
curve of the star CO Ori (Rostopchina et al. 2007),
which is a long series of photometric observations
of this star folded with a period of $12.4$ yr. The behavior
of the linear polarization in this star (observed
synchronously with photometry) strongly suggests
that periodic circumstellar extinction variations are
responsible for the cyclic variations in its brightness.
We see from Fig. 8 that the cyclic brightness variations
in CO Ori are generally similar in shape to the
model light curve presented in Fig. 2. It should be
noted that similar (in shape) light curves can also be
obtained in models with noticeably differing parameters.
Therefore, the shapes of the observed activity
cycles and their amplitudes cannot be used, for example,
to estimate the mass of the secondary component
mass.

In conclusion, it should be noted that the accretion
disk of the primary component, if it contains dust, can
also be the source of photometric activity of a young
binary system. This requires that the disk radius be
larger than the radius of the dust sublimation zone.
For Herbig Ae stars, to which most of the UX Ori
stars belong, the radius of this zone is $0.5$ AU.
Therefore, at a disk radius of the order of several
AU, the existence of circumstellar extinction and its
possible variations should be taken into account when
the light curves of young binary systems are analyzed.
\\
\section{ACKNOWLEDGMENTS}
We are grateful to the referee for helpful remarks.
This work was supported by the "Origin and Evolution
of Stars and Galaxies" Program of the Presidium
of the Russian Academy of Sciences and the
Program for Support of Leading Scientific Schools
(NSh-1318.2008.2 and NSh-6110.2008.2).
\clearpage
\begin{center}
{\Large\bf REFERENCES}
\end{center}
1. S. A. Artemenko, K. N. Grankin, and P. P. Petrov, Astron. Zh. 73, 186 (2010) [Phys. At. Nucl. 73, 163 (2010)].\\
2. T. Birnstiel, C. P. Dullemond, and F. Brauer, Astron. Astrophys. 503, L5 (2009). \\
3. V. P. Grinin, E. Stempels, G. Gahm, et al., Astron. Astrophys. 489, 1233 (2008).\\
4. T. V. Demidova, N. Ya. Sotnikova, and V. P. Grinin, Pis'ma Astron. Zh. 36, 445 (2010a) [Astron. Lett. 36, 422 (2010)].\\
5. R. Garcia Lopez, A. Natta, L. Testi, and E. Habart, Astron. Astrophys. 459, 837 (2006).\\
6. J. Muzerolle, P. D'Alessio, N. Calvet, and L. Hartmann, Astrophys. J. 617, 406 (2004).\\
7. A.Natta and B.Whitney, Astron. Astrophys. 364, 633 (2000).\\
8. A. N. Rostopchina, Astron. Zh. 76, 136 (1999) [Astron. Rep. 43, 113 (1999)].\\
9. A. N. Rostopchina, D. N. Shakhovskoi, V. P. Grinin, and A. A. Lomach, Astron. Zh. 84, 60 (2007) [Astron. Rep. 51, 55 (2007)].\\
10. N. Ya. Sotnikova, Astrofizika 39, 259 (1996).\\
11. N. Ya. Sotnikova and V. P. Grinin, Pis'ma Astron. Zh. 33, 667 (2007) [Astron. Lett. 33, 594 (2007)].\\
12. L. V. Tambovtseva, V. P. Grinin, B. Rodgers, and O. Kozlova, Astron. Zh. 78, 514 (2001) [Astron. Rep. 45, 442 (2001)].\\
13. D. N. Shakhovskoi, V. P. Grinin, and A. N. Rostopchina, Astrofizika 48, 166 (2005).\\

Translated by G. Rudnitskii

\clearpage
\begin{figure}[!h]\begin{center}
  \makebox[0.8\textwidth]{\includegraphics[scale=0.8]{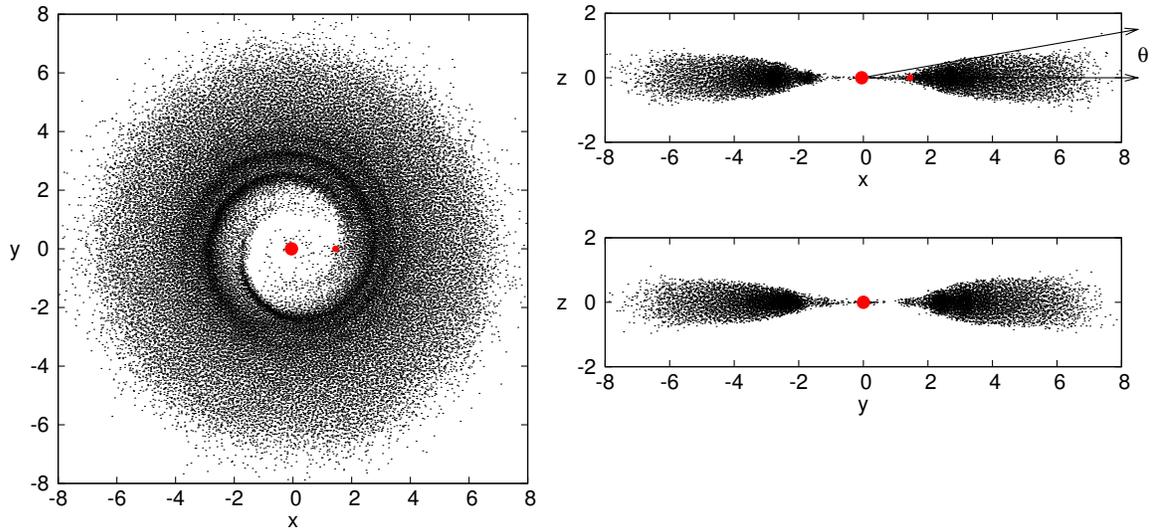}
  }
 \caption{Distribution of matter in the binary model 3 ($e = 0.5$, $q = 0.03$): (a) top view; (b) and (c) disk sections in the $xz$ and $yz$
planes. The scale along all three axes is given in units of the orbital semimajor axis.}
 \label{disk}
\end{center}
\end{figure}
\begin{figure}[!h]\begin{center}
  \makebox[0.6\textwidth]{\includegraphics[scale=1.2]{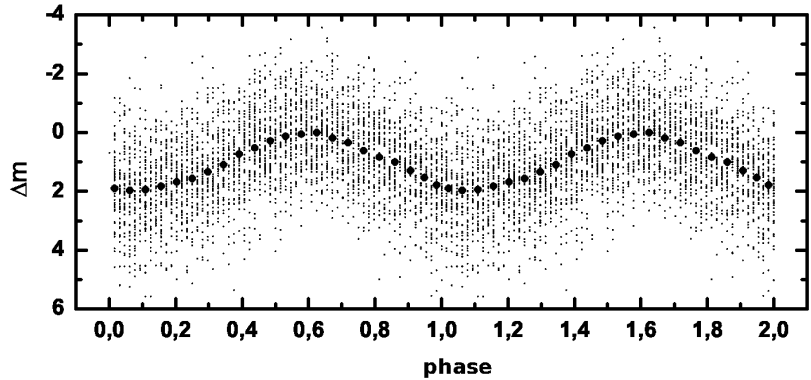}
  }
 \caption{Phase light curve for model 1; the inclination of the binary orbit to the line of sight is  $\theta = 7^{\circ}.5$ , the rotation angle of the
apsidal line is $180^{\circ}$. The dots indicate the nonaveraged values of $\Delta m$; the circles indicates the averaged and smoothed values.}
 \label{phase}
\end{center}
\end{figure}
\begin{figure}[!h]\begin{center}
  \makebox[0.6\textwidth]{ \includegraphics[scale=0.9]{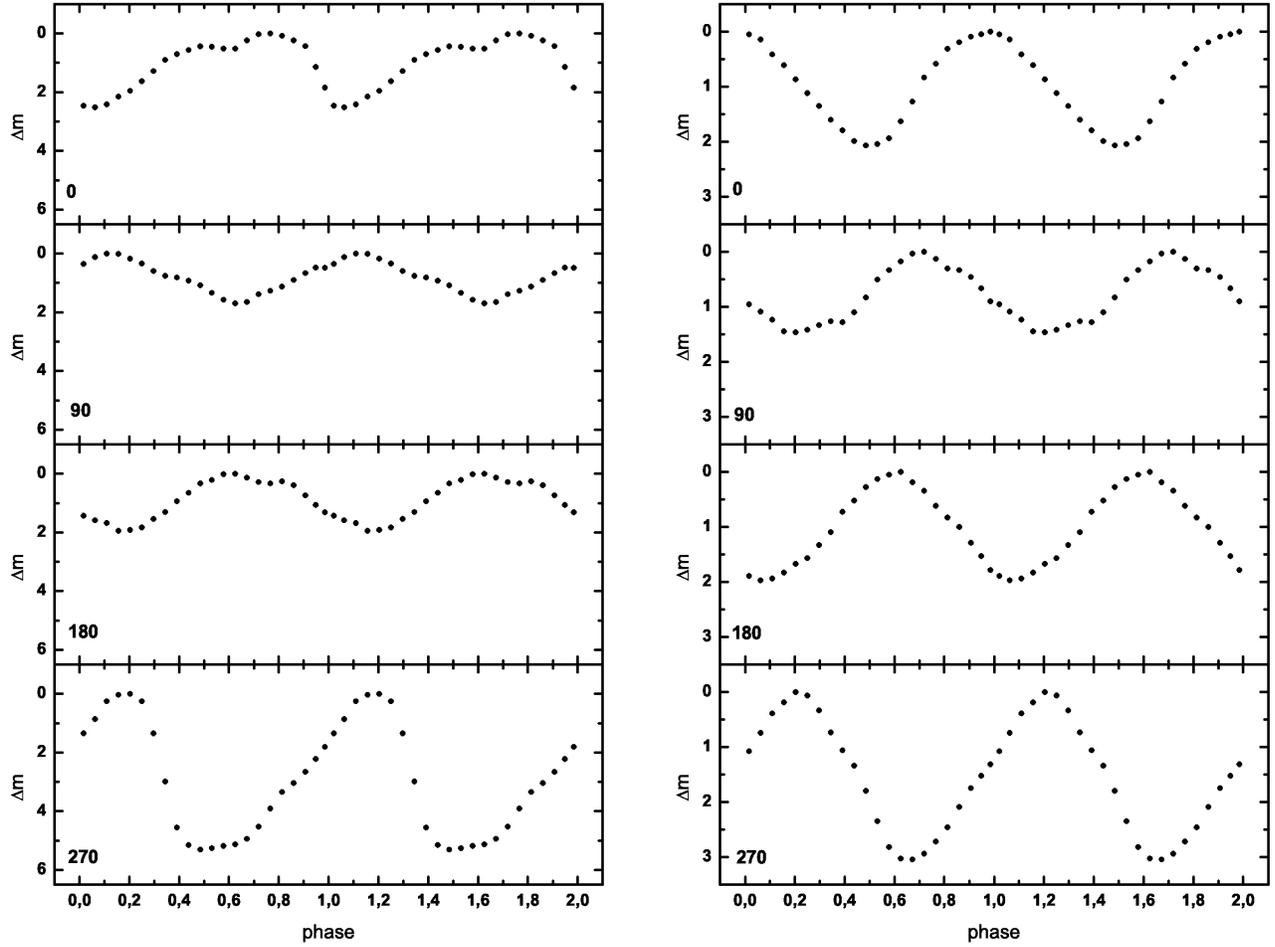}
  }
 \caption{Phase light curves in model 1 ($e = 0.3$, $q = 0.1$): (a) the line of sight lies in the orbital plane, (b) the disk is inclined at
an angle of $7^{\circ}.5$ to the line of sight. The rotation angle of the apsidal line relative to the observer is indicated in the lower left
corner of each diagram.} \label{model1}
\end{center}
\end{figure}
\newpage
\begin{figure}[!h]\begin{center}
  \makebox[0.6\textwidth]{\includegraphics[scale=0.9]{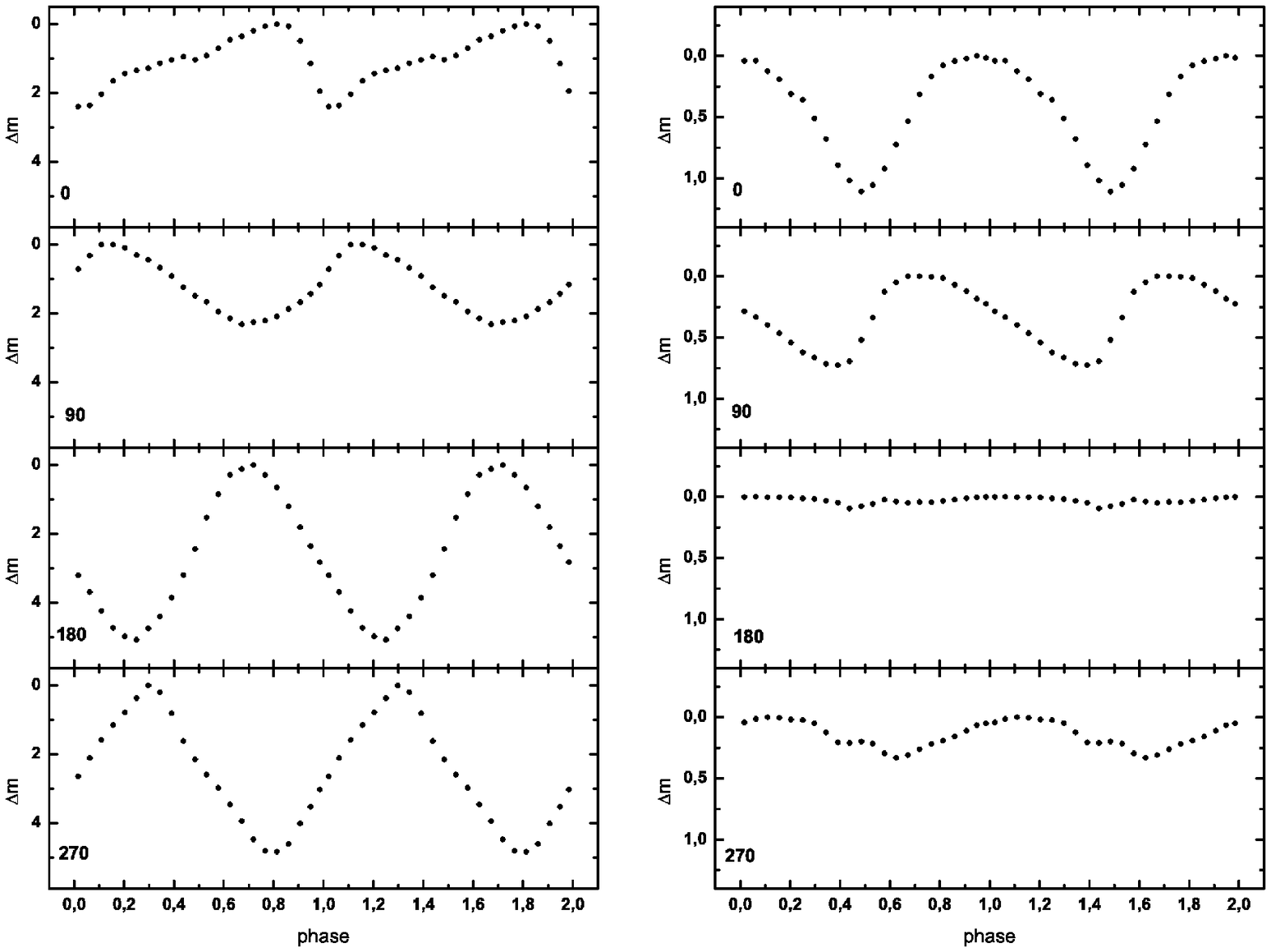}
  }
 \caption{Same as Fig. 3 for model 2 ($e = 0.5$, $q = 0.1$); (b) the binary orbit is inclined at an angle of $12^{\circ}$ to the line of sight.} 
\label{model2}
\end{center}
\end{figure}

\newpage
\begin{figure}[!h]\begin{center}
  \makebox[0.6\textwidth]{\includegraphics[scale=0.9]{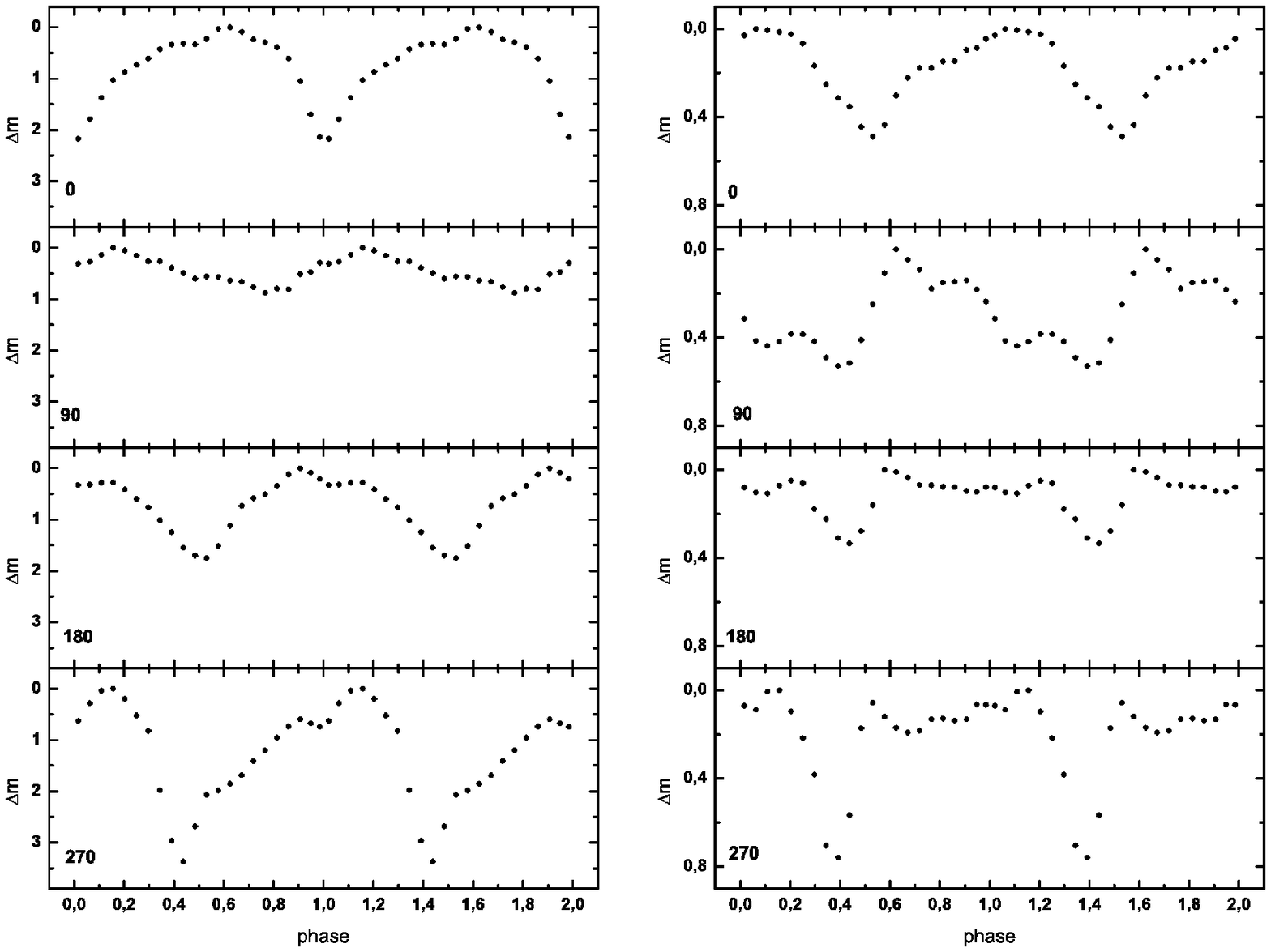}
  }
 \caption{Same as Fig. 3 for model 3 ($e = 0.5$, $q = 0.03$); (b) the binary orbit is inclined at an angle of $10^{\circ}$ 
to the line of sight.}\label{model3}
\end{center}
\end{figure}
\newpage
\begin{figure}[!h]\begin{center}
  \makebox[0.6\textwidth]{\includegraphics[scale=0.9]{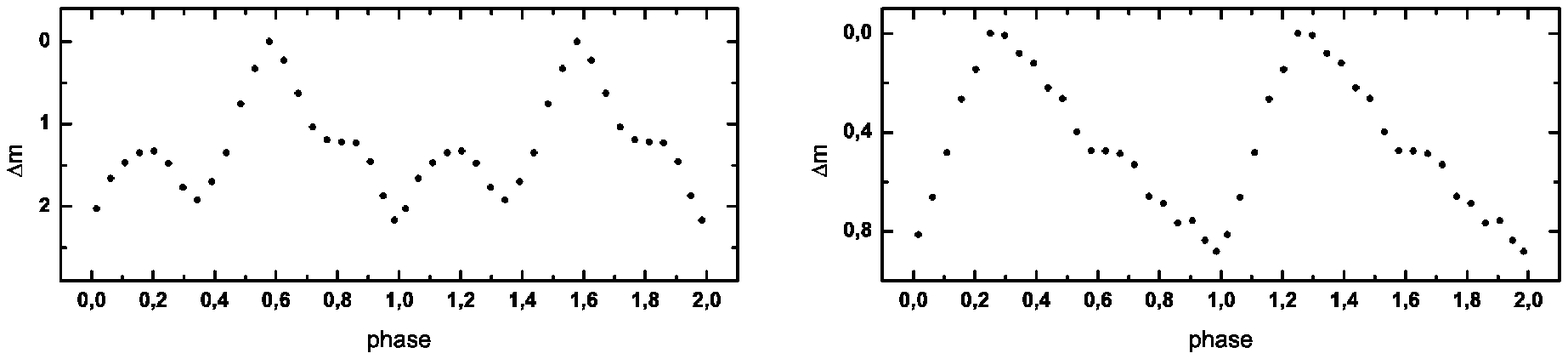}
  }
 \caption{Same as Fig. 3 for model 4 ($e = 0.0$, $q = 0.01$); (b) the binary orbit is inclined at an angle of $9^{\circ}$ 
to the line of sight.}\label{model4}
\end{center}
\end{figure}

\begin{figure}[!h]\begin{center}
  \makebox[0.6\textwidth]{\includegraphics[scale=0.9]{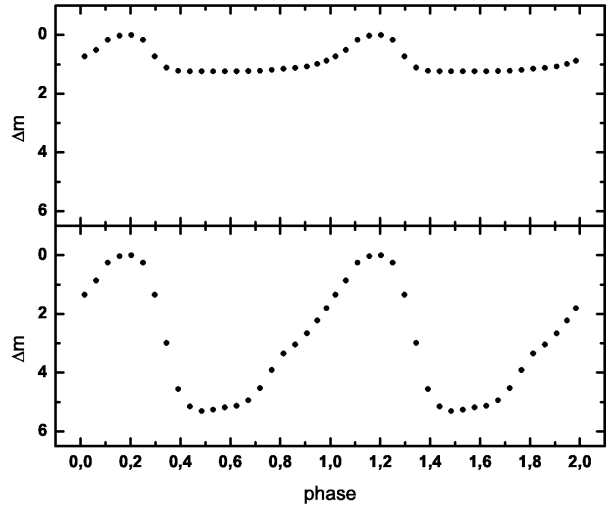}
  }
 \caption{Diagram illustrating the influence of scattered radiation ($I_{sc} = 0.1I_{\ast}$) on the behavior of the binary brightness; the light
curve in model 1 (a) with and (b) without scattered light.}
 \label{Isc}
\end{center}
\end{figure}

\begin{figure}[!h]\begin{center}
  \makebox[0.6\textwidth]{\includegraphics[scale=0.5]{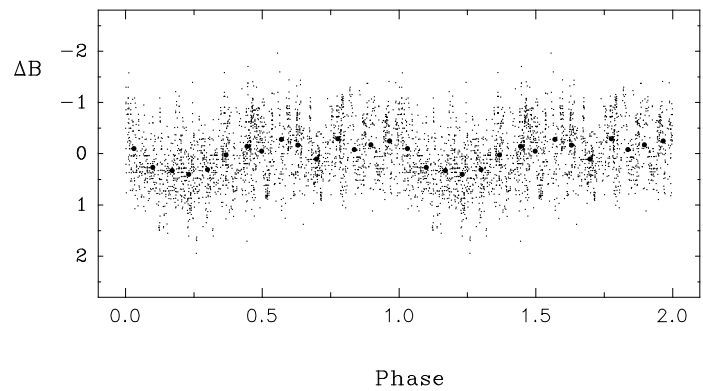}
  }
 \caption{$B$-band light curve of CO Ori folded with a period of 12.4 yr based on data from Rostopchina et al. (2007).}
 \label{co}
\end{center}
\end{figure}

\end{document}